# Scarce or Missed Opportunities?

# Structural Peers and Latin America's Diversification Frontier

Saangkeub Lee[a], Dawoon Jeong [b,c*], Wonsub Eum[d], Jeong-Dong Lee[a]

[a] *Technology Management, Economics and Policy Program, Seoul National University, 1 Gwanak-ro, Gwanak-gu, Seoul, 08826, Republic of Korea*

[b] *Department of Sociology, University of Chicago, Chicago, IL, 60637, USA*

[c] *Knowledge Lab, University of Chicago, Chicago, IL, 60637, USA*

[d] *Department of Global Business, Rikkyo University, 3 Chome-34-1 Nishiikebukuro, Toshima City, Tokyo, 171-8501, Japan*

** (Corresponding authors: jdwoon0523@uchicago.edu)*

**Abstract**

Latin America's limited export upgrading is often attributed to failures to seize available diversification opportunities. We distinguish two possibilities: countries failed to realize opportunities demonstrated by structurally similar economies, or upgrading-oriented options were scarce within this peer-revealed set. Using export data for 105 countries from 1962 to 2024, we construct country-specific opportunity sets with an imitation-density measure associated with subsequent product entry. Latin America's deficit arose mainly from scarce upgrading-oriented candidates rather than weak aggregate conversion. Conversion rates matched the global benchmark, but regional frontiers contained few manufacturing products and almost no higher-complexity manufacturing candidates and, unlike East and Southeast Asia, did not shift toward them over time. An illustrative decomposition associates most of the regional upgrading-entry gap with candidate availability. The findings imply that industrial policy should expand feasible upgrading pathways rather than focus only on converting existing opportunities.

## 1. Introduction

Why has Latin America's export diversification contributed so little to industrial upgrading? The most common diagnosis, implicit in much of the policy debate, is one of missed opportunities: the region possessed feasible pathways toward more sophisticated manufacturing but failed to seize them, whether because of macroeconomic volatility, weak industrial policy, premature liberalization or institutional shortcomings. This diagnosis carries a strong implication—that the opportunities were there, and that better policy execution would have converted them into structural change.

This article examines a prior question that the missed-opportunities diagnosis takes for granted: what did the opportunity set actually contain? We separate the problem into two logically distinct hypotheses. Under the first, upgrading opportunities were available but under-realized: Latin American countries systematically failed to convert feasible diversification options into actual comparative advantage. Under the second, the opportunities themselves were scarce: the set of products realistically within reach at each point in time contained few manufacturing products and fewer still of above-average complexity, so that even a normal rate of realization could not have produced substantial upgrading. The two hypotheses call for very different policy responses—the first points to implementation and coordination failures downstream, the second to the structural reproduction of a shallow opportunity set upstream—yet the existing literature rarely distinguishes them empirically, largely because the feasible opportunity set of a country at a given moment is not directly observable.

Our approach is to make one well-defined component of that set observable, by drawing on the experience of structurally similar economies. Throughout the article we refer to this component as the peer-revealed opportunity set: it is the set of options revealed by the observed product composition of a country's contemporaneous structural peers, not the universe of everything a country could conceivably enter through foreign investment, strategic industrial policy, technology licensing or path-defying diversification. The intuition follows a long tradition of comparator-based reasoning in development economics (Lin and Monga, 2011; Lin, 2012; Lin and Fu, 2024) and connects with evidence that knowledge about export capabilities diffuses between economies with similar productive structures (Bahar, Hausmann and Hidalgo, 2014). It resonates, too, with the catch-up literature, in which the imitation of economies in comparable positions has historically been a central strategy of capability building (Abramovitz, 1986; Kim, 1997; Lee, 2013). If countries whose export baskets closely resemble that of a focal country already hold comparative advantage in a product the focal country does not yet export competitively, that product is a plausible member of the focal country's peer-revealed set: the capabilities it requires are demonstrably compatible with an export structure like the focal country's own. We operationalize this idea through an *imitation density* measure (Jeong, Lee and Eum, 2024), defined for each country, product and year as the export-similarity-weighted share of a country's structural peers that hold revealed comparative advantage in a candidate product. Ranking each country-year's candidate products by imitation density and taking the twenty top-ranked products within that country-year defines what we call the peer-revealed priority set (for brevity, the high-imitation frontier).

Three features of this construction matter for our question. First, the frontier is disciplined by predictive validity. Before using the frontier diagnostically, we assess whether imitation density is systematically associated with subsequent diversification. When candidate products are ranked within each country-year, the ten percent with the highest imitation density enter within ten years at a rate of 26.3%, compared with 4.9% for the ten percent with the lowest imitation density. In the common regression sample of 2,904,005 country-product-year observations, this positive association remains statistically significant after controlling for relatedness density and product characteristics and including country-year and product fixed effects. The result also remains robust in an extended specification with product-year fixed effects.

Second, the frontier is country-specific and time-varying: as a country's export structure evolves, so do its structural peers and hence its revealed opportunities. Third, and crucially for our purposes, the frontier has measurable content: we can ask not only whether frontier products were subsequently entered, but what share of them were manufacturing products, how complex they were, and how that composition evolved over six decades.

Within the peer-revealed frontier, the answer to our title question is primarily scarcity. Latin American countries showed no large descriptive conversion shortfall relative to the global benchmark: 30.7% of frontier products entered within ten years, matching the global benchmark rate of 30.7%, while sustained-entry rates were also close (13.7% compared with 14.6%). Where Latin America diverged sharply was in what the frontier contained. Manufacturing products accounted for 33.2% of the region's frontier, against 41.8% globally and 47.2% in East and Southeast Asia; manufacturing products of above-average complexity accounted for just 2.4%, against 13.4% globally and 16.8% in East and Southeast Asia. On average, a Latin American country-year had 0.49 higher-complexity manufacturing candidates in its frontier—one such option roughly every two years—compared with 3.43 in East and Southeast Asia, and roughly five out of six Latin American country-years had none at all. Nor was this a static difference: over time, the East and Southeast Asian frontier shifted decisively toward manufacturing and, from the late 1990s, toward higher-complexity manufacturing, while the Latin American frontier's manufacturing content rose in the 1970s–1980s but receded after 2000, a period coinciding with the commodity super-cycle. An illustrative decomposition of the observed regional upgrading-entry gap indicates that roughly four fifths of the gap is associated with differences in candidate availability rather than conversion.

The scarcity finding does not fully exonerate realization. The few higher-complexity manufacturing candidates that did appear in Latin American frontiers had a lower pooled entry rate than in the global benchmark (21.2% versus 32.9%), although the small number of such observations counsels caution in interpreting this descriptive difference. The more robust conclusion is compositional, and it is deliberately bounded: within the set of options validated by the experience of contemporaneous structural peers, the region's problem was less that it ignored upgrading options than that its peers—economies with similar productive structures—had little upgrading experience to reveal. This reading connects directly with the structuralist tradition associated with the Economic Commission for Latin America and the Caribbean (ECLAC): a productive structure specialized in low-complexity activities reproduces itself not only internally, through the paucity of related capabilities, but also externally, through the composition of the peer experience available for reference (Prebisch, 1950; Ocampo, 2005; ECLAC, 2012; Cimoli and Porcile, 2014).

The article makes three contributions. First, it provides a tractable empirical distinction between how many higher-complexity manufacturing candidates appear in the peer-revealed frontier and how often those candidates subsequently enter. An illustrative decomposition shows that, in the benchmark comparison, differences in the number of such candidates account for most of the observed regional gap in upgrading entries. Second, building on and extending earlier work on peer-based diversification signals (Jeong, Lee and Eum, 2024), it deploys imitation density as a diagnostic complement to the product-space relatedness measures now widely used in diversification analysis (Hidalgo et al., 2007; Hausmann and Klinger, 2007): the two measures overlap in only about half of their top-ranked candidates, and products flagged only by peer experience enter at more than twice the rate of products flagged by neither, so peer benchmarking adds genuine information to internal-capability screening. Third, it offers a caution for opportunity-identification exercises in industrial policy: a screening tool that predicts entry well may nonetheless surface a set of options that does little for structural change, so the composition of the identified set deserves the same scrutiny as its predictive power.

The remainder of the article proceeds as follows. Section 2 situates the approach between internal-capability and external-reference perspectives on diversification and explains why the composition of the opportunity set matters for structural change. Section 3 describes the data and the construction of export similarity, imitation density, entry outcomes, the peer-revealed frontier and the relatedness benchmark. Section 4 presents the results in the order of the argument: validation, aggregate conversion, composition, the availability–realization decomposition, evolution and intra-regional heterogeneity, and the complementarity between peer experience and product relatedness. Section 5 discusses implications for the interpretation of Latin America's diversification record and for policy, and section 6 concludes.

## 2. Literature review

### 2.1. Internal capability: the relatedness perspective

The dominant empirical approach to diversification possibilities starts from a country's own current portfolio. The product-space literature showed that countries rarely diversify in arbitrary directions; they move predominantly into products that are "close," in terms of required capabilities, to those they already export competitively (Hidalgo et al., 2007; Hausmann and Klinger, 2007; Hidalgo and Hausmann, 2009). Because new export activities require bundles of capabilities—technology, worker skills, input supply chains, quality management, market access, organizational routines—candidates that share such requirements with the existing basket are more attainable. The insight has been generalized as the *principle* *of relatedness*, which holds across products, industries, technologies and occupations (Boschma, 2017; Hidalgo et al., 2018), and relatedness or density measures are now standard predictors of new revealed comparative advantage (Coniglio et al., 2021; Pinheiro et al., 2022).

For a developing region, however, an exclusively internal perspective has a well-known limitation, one that structuralist thought anticipated long before the product space was mapped: when the existing basket is concentrated in low-complexity, weakly connected products, the candidates that relatedness identifies as attainable may themselves offer limited upgrading potential. Indeed, as the complexity of a target product rises, its relatedness to a low-complexity export basket declines almost by construction, progressively narrowing the scope of path-dependent upgrading (Jeong, Lee and Eum, 2024); the smart-specialization prescription of targeting products that are simultaneously related and complex (Balland et al., 2019) then approaches an empty set. The relatedness perspective answers the question “what is close to what I already do?”; it cannot, by construction, tell a country what economies in comparable positions have found to be within reach. That is the role of an external reference.

## 2.2. External reference: structurally similar economies and imitation density

We define a country’s structural peers not by geography, income group or institutional category, but by the similarity of export structures at each point in time. This choice follows the product-space premise that an export basket is an empirical window onto underlying productive capabilities (Hidalgo et al., 2007; Hidalgo and Hausmann, 2009): two countries with similar revealed specialization patterns plausibly share elements of a capability profile, whatever their location or income. It also gives empirical content to the comparator logic of New Structural Economics, which recommends that developing countries scan the experience of economies in comparable positions for latent upgrading directions (Lin and Monga, 2011; Lin, 2012), and it aligns with evidence that comparative advantage diffuses more readily between similar economies (Bahar, Hausmann and Hidalgo, 2014). Evidence at the regional and country levels points the same way: neighboring regions' industry structures predict the emergence of new local industries (Boschma, Martín and Minondo, 2017), spatial spillovers accelerate export diversification (Basile, Parteka and Pittiglio, 2018), and structural relatedness between trading partners shapes diversification even conditional on geographic proximity (Jun et al., 2020). Whereas geographic proximity confines the reference set to neighbors, structural similarity can connect a country to relevant benchmarks anywhere in the world. The catch-up literature supplies the underlying rationale: development for technologically lagging economies is largely a process of converging toward capabilities demonstrated elsewhere (Abramovitz, 1986; Kim, 1997; Lee, 2013), and the most informative benchmarks are economies sufficiently similar in endowments—neither so distant that their path is non-replicable nor so close that they reveal nothing new (Lin and Fu, 2024). A related strand of applied work identifies export candidates by collaborative-filtering methods over the country-product matrix; relative to that strand, our contribution lies in an interpretable peer-density measure, the separation of external peer signals from internal relatedness, and—centrally—the distinction between the predictability of candidates and their developmental composition.

Given a set of structural peers, imitation density measures, for each candidate product a country does not yet export competitively, how intensively that product is held across the peer group, weighting each peer by its similarity to the focal country (Jeong, Lee and Eum, 2024). A high value means that the product recurs in the portfolios of economies whose revealed capability profiles resemble the focal country's own—economies for which the technological, input, quality, market-access and institutional requirements of the product have demonstrably been compatible with a similar structure. The measure is deliberately agnostic about transmission mechanisms: policy emulation, firm-level learning, foreign direct investment, supply-chain linkages and market information can all make peer experience relevant, and we do not attempt to identify them separately. What the measure summarizes is the density of external experience surrounding each candidate option. In the companion study, imitation density is positively and significantly associated with subsequent entry even conditional on relatedness density, the two signals interact synergistically, and the peer signal is most strongly associated with entry among low-complexity countries and for products of low-to-intermediate complexity—consistent with a bridging role in the transition from simple to complex products (Jeong, Lee and Eum, 2024). That last regularity already intimates the concern at the center of this article: the products around which peer experience is densest need not be the products through which structural change proceeds.

Three terminological cautions apply throughout. First, "imitation" refers to the use of observed peer portfolios as reference information—peer-revealed empirical feasibility—not to any claim that countries deliberately copy one another. Second, imitation density is one signal of feasibility, not a complete measurement of it: products absent from peer portfolios may still be attainable through foreign investment, strategic policy or technology transfer, and products present in peer portfolios may be unsuitable for a focal country's endowments or market size. All claims in this article therefore concern the peer-revealed opportunity set, and we phrase them accordingly. Third, "entry" in this article denotes the activation of revealed comparative advantage (RCA), including reactivation of products inactive for at least five years; it should not be read as first-ever industry entry. Entry so defined is an extensive-margin notion—the addition of new competitively exported lines—distinct from changes in the dispersion or concentration of the export basket measured by entropy-type indices (Imbs and Wacziarg, 2003; Cadot, Carrère and Strauss-Kahn, 2011).

## 2.3. Why composition matters: opportunities and structural change

Whether an opportunity-identification device serves development analysis depends not only on whether the products it flags are subsequently entered, but on what kind of products they are. A long literature cautions against equating diversification with development: what is exported matters for subsequent growth (Hausmann, Hwang and Rodrik, 2007; Lall, Weiss and Zhang, 2006), diversification proceeds chiefly at the extensive margin through export discoveries whose emergence is shaped by entry barriers and market failures (Klinger and Lederman, 2006 and 2011; Cadot, Carrère and Strauss-Kahn, 2011), and the quality composition of the basket is a developmental dimension in its own right (Henn, Papageorgiou and Spatafora, 2013). Manufacturing retains a particular place in this discussion because of its association with productivity growth, learning and linkages, even as premature deindustrialization narrows the manufacturing window for late developers (Rodrik, 2016) and manufacturing activity concentrates in fewer countries (Haraguchi, Cheng and Smeets, 2017).

The ECLAC structuralist tradition sharpens the point for Latin America: development requires progressive structural change—a sustained reallocation toward activities of higher productivity, knowledge intensity and demand dynamism—not merely the addition of export lines (Prebisch, 1950; Ocampo, 2005; ECLAC, 2012; Cimoli and Porcile, 2014). From this standpoint, the central diagnostic question about any opportunity set is compositional: does it contain the manufacturing and higher-complexity options through which progressive structural change proceeds, and does it come to contain more of them over time?

We therefore evaluate the peer-revealed frontier on two margins. Availability asks how many manufacturing and higher-complexity manufacturing candidates the frontier contained at each moment. Realization asks what fraction of frontier candidates—overall and within the higher-complexity manufacturing subset—were converted into actual entry. Keeping the two margins separate, and formally decomposing the regional upgrading gap into them, is what allows the title question to be answered empirically rather than assumed.

## 3. Data and Methods

### 3.1. Data

The analysis uses country-product-year export data from the Atlas of Economic Complexity (Growth Lab at Harvard University, 2026), covering 1962–2024. The estimation sample comprises 105 countries and 786 products at the four-digit level of the Standard International Trade Classification (SITC), Rev 2; the 786 products are those observed in the Atlas SITC Rev 2 four-digit data over the analysis period and usable for the constructions below.

A country $c$ is active in product $p$ in year $t$ when its revealed comparative advantage exceeds unity, $RCA(c',p,t) > 1$, with RCA defined in the standard Balassa (1965) form as the product's share in the country's exports relative to its share in world exports ( $RCA(c,p) = (X_{cp}/\sum_{p'} X_{cp'})/(\sum_{c'} X_{c'p}/\sum_{c'p'} X_{c'p'}$ ). Products with $RCA(c',p,t) \leq 1$ constitute the country-year's candidate set—products not yet exported competitively, over which diversification outcomes are subsequently evaluated. When a country-year is observed but a product's export value is not reported, RCA is set to zero; country-years missing entirely from the data are dropped. Candidates are defined over the full four-digit product space rather than manufacturing alone, since peer portfolios span the whole export structure; manufacturing status and complexity are then measured as characteristics of the identified candidates.

Entry occurs when a candidate product in base year t becomes active at least once during the following ten years (t+1 to t+10). Because the data end in 2024, base years for the ten-year outcome run through 2014. To limit the influence of transitory RCA fluctuations, we also compute sustained entry, defined as the product being active in three consecutive years within the ten-year window. Two sample restrictions apply to the baseline candidate observations: the full ten-year window must be observable, and products that were active at any point in the five years preceding the base year are excluded, so that short-run re-entries of recently lapsed products are not counted as new diversification.

### 3.2. Export similarity, peer weights and imitation density

Structural similarity between countries is measured with the export-similarity index of Bahar, Hausmann and Hidalgo (2014), adopted here without modification. For each country, product and year, define the log-transformed RCA

$$r_{cpt} = \ln\left(RCA_{cp} + \varepsilon\right)$$

$$\varepsilon = 0.1$$

where the constant keeps zero-RCA products in the vector and the log transformation prevents a few very high RCA values from dominating cross-country correlations. Export similarity between countries c and c′ in year t is the Pearson correlation of their log-RCA vectors across the 786 products:

$$S_{cc't} = \frac{\sum_p (r_{cpt} - \overline{r_{ct}})(r_{c'pt} - \overline{r_{c't}})}{\sqrt{\sum_p \left(r_{cpt} - \overline{r_{ct}}\right)^2 \left(r_{c'pt} - \overline{r_{c't}}\right)^2}}$$

Similarity is computed within the 105-country estimation sample and recalculated every year, so each country's peer group is dynamic: as an export structure changes, so does the set of economies that resemble it. Negative correlations indicate opposed rather than similar structures and are not treated as structural similarity; we retain only $S^{+}_{cc't} = \max(S_{cc't}, 0)$ and normalize positive similarities to sum to one within each country-year, yielding peer weights $\omega_{cc't} = \frac{S^{+}_{cc't}}{\sum_{c' \neq c} S^{+}_{cc't}}$ with the focal country excluded. In the rare country-years with no positively similar peer (3 of 6,579 country-year pairs for which similarity is computable), no reference group is defined, and imitation density is set to zero for all candidates. Imitation density for candidate product $p$ of country $c$ in year $t$ is then the peer-weighted share of structural peers active in the product:

$$\omega^{imi}_{cpt} = \frac{\sum_{c' \neq c} M_{c'pt} S^{+}_{cc't}}{\sum_{c' \neq c} S^{+}_{cc't}} = \sum_{c' \neq c} \omega_{cc't} M_{c'pt}$$

where $M_{c'pt}$ equals one when $RCA(c', p, t) > 1$. The measure lies between zero and one and is used only for candidate products. A high value requires jointly that peers with similar revealed capability profiles exist, that many of them hold the product, and that the holders are precisely the structurally closest economies rather than arbitrary countries. Imitation density is calculated using the same procedure as in Jeong, Lee and Eum (2024).

### 3.3. The peer-revealed frontier and product characteristics

Imitation density enters the analysis in two ways. First, candidates are ranked within each country-year and grouped into deciles, so that entry rates can be compared between candidates with relatively weak and relatively strong peer signals in the same country and year; the design deliberately avoids pooling across country-years. Second, the top-ranked candidates in each country-year define the peer-revealed priority set, or high-imitation frontier—the options most strongly revealed by peer experience at that moment. "Frontier" here denotes a priority screening set, not a technological frontier or an optimal list. The baseline frontier comprises the twenty top-ranked candidates in each country-year. The top-twenty criterion is a relative, within-country-year rank rather than an absolute threshold: in country-years with fewer than twenty candidates carrying a positive imitation density, products with zero imitation density can enter the set, in which case the frontier should be read as the relatively top-ranked candidates of that country-year rather than as a set of high absolute density.

Frontier composition is measured with two product characteristics. Manufacturing status assigns Mfg(p) = 1 to four-digit products in SITC two-digit divisions 51–59, 61–69, 71–79, 81–85 and 87–89. Product complexity uses the Product Complexity Index (PCI) of Hidalgo and Hausmann (2009) at the product-year level, standardized within each year across the products with observed PCI, so that a value of zero denotes average complexity in that year's product distribution. Higher-complexity manufacturing candidates are then defined operationally as candidates with Mfg(p) = 1 and year-standardized PCI $\geq 0$. Composition shares are computed over simple product counts (unweighted by density); decile compositions aggregate candidate-product observations within deciles, while regional and period comparisons of the frontier first compute each country-year's composition and then average across country-years, so the unit of aggregation is the country-year.

### 3.4. The relatedness benchmark

To establish what peer experience adds beyond internal capability, we compute the standard relatedness density benchmark. Product-to-product proximity in year t is the minimum of the two conditional probabilities of co-occurring comparative advantage across countries (Hidalgo et al., 2007), computed—unlike export similarity—over the full set of countries with usable RCA information in the Atlas for that year, so that the global product-space structure is not conditioned on the estimation sample; proximity is recalculated annually. Relatedness density for candidate $p$ of country $c$ in year $t$ is the proximity-weighted share of the country's active products among all products, $\omega_{cpt}^{rel} = \frac{\sum_{q \neq p} \phi_{pq} M_{cpt}}{\sum_{q \neq p} \phi_{pq}}$·, assigned to the same baseline candidate observations as imitation density. A top-ranked relatedness set is constructed within each country-year exactly as for imitation density, and candidates are cross-classified into four groups: top-ranked in both measures, in imitation density only, in relatedness density only, or in neither. The relatedness comparison covers baseline candidate observations for 1965–2014, matching the main comparison window.

## 4. Empirical Results

### 4.1. Peer experience identifies feasible opportunities

The diagnostic use of the peer-revealed frontier presupposes that imitation density is systematically related to future diversification; this subsection establishes that it is.

[Insert Figure 1 Here]

Figure 1 ranks each country-year's candidates into imitation-density deciles and plots the share entering within ten years. In the global sample, entry rises monotonically from 4.9% in the bottom decile to 26.3% in the top decile—a more than fivefold difference between candidates with weak and strong peer signals in the same country and year. The gradient is not a feature of any single region: in Latin America the corresponding rates are 4.4% and 26.8%, and the same pattern holds in East and Southeast Asia and in Africa.

[Insert Figure 2 Here]

Figure 2 shows that the relationship is not an artifact of transitory RCA crossings: under the stricter sustained-entry criterion, the rate rises from 1.1% in the bottom decile to 11.9% in the top decile globally (0.9% to 11.4% in Latin America). Table 1 confirms the relationship in a linear probability framework at the country-product-year level, with country-year fixed effects throughout, so that identification comes from comparisons among candidates of the same country in the same year.

[Insert Table 1 Here]

The imitation-density coefficient is positive and significant in every specification. Conditioning on relatedness density preserves roughly two-thirds of the coefficient (0.047 against 0.068), and it is essentially unchanged by product characteristics or product fixed effects; in the fullest specification, a 0.10 increase in imitation density is associated with an entry probability higher by about 4.7 percentage points. Extended specifications—including candidate history, product ubiquity, two-digit-sector-by-year fixed effects and product-by-year fixed effects—leave the conclusion intact (results available from the authors upon request).

[Insert Table 2 Here]

Table 2 shows that the same conclusion holds for sustained entry. Peer experience, in short, carries real information about what is within reach—in Latin America no less than elsewhere. The question is what that information contained.

## 4.2. Aggregate conversion: Latin America realized its frontier at the benchmark rate

Table 3 addresses the realization hypothesis at the level of the frontier. Among the frontier candidates identified each country-year, the share entering within ten years was 30.7% in Latin America, against 30.7% in the global benchmark and 31.9% in East and Southeast Asia; sustained entry was 13.7% against 14.6% globally.

[Insert Table 3 Here]

These figures are descriptive comparisons of country-year averages rather than formal tests of equality, and the global benchmark includes Latin America itself. They should therefore be read as indicating that Latin America's aggregate frontier-conversion rate was close to the global benchmark~~,~~ rather than as establishing exact equality. Two conclusions should be kept separate. For the frontier as a whole, the descriptive results do not indicate a large Latin American conversion shortfall. For the upgrading-oriented subset of the frontier, section 4.4 documents a lower Latin American conversion rate, imprecisely estimated because such candidates were rare. The claim defended in this article is therefore deliberately bounded: within the peer-revealed frontier, the descriptive comparison does not indicate a large aggregate realization gap~~,~~ while a smaller and less precisely estimated realization gap may remain among the rare higher-complexity manufacturing candidates.

## 4.3. Availability: what the frontier contained

The remaining columns of table 3 locate the divergence. Manufacturing products made up 33.2% of the Latin American frontier, well below the global 41.8% and the East and Southeast Asian 47.2%. The gap widens dramatically once complexity is considered: manufacturing products of above-average complexity accounted for 2.4% of the Latin American frontier, against 13.4% globally and 16.8% in East and Southeast Asia—a sevenfold difference with the Asian comparison. The frontier's mean standardized PCI was correspondingly lower in Latin America (−1.38 against −1.11 globally and −1.02 in East and Southeast Asia). Only Africa shows a thinner upgrading content.

[Insert Figure 3 Here]

Because a compositional gradient of this kind is partly built into the measure—products widely held among any country's peers tend to be widely held generally, and ubiquitous products have low complexity almost by construction—the level of frontier complexity is uninformative on its own. Indeed, across all candidates, average standardized PCI falls with the imitation-density decile—from +0.60 in the bottom decile to −0.92 in the top decile globally—and the share of above-average-complexity manufacturing falls from 79.2% to 15.0%.

[Insert Figure 4 Here]

[Insert Figure 5 Here]

What matters for the regional diagnosis is therefore the differences across regions facing the same construction: in the top decile, above-average-complexity manufacturing still made up 22.0% of East and Southeast Asian candidates but only 5.6% of Latin American ones. Manufacturing status and complexity are related but far from interchangeable characteristics: manufacturing products are more likely to carry above-average PCI, but the distributions overlap substantially, which is why the analysis tracks both and their intersection.

## 4.4. The availability–realization decomposition

The title question can be answered quantitatively. The expected number of upgrading entries per country-year is the product of two margins:

$$E[upgrading\ entries] = E[Number\ of\ higher\ complexity\ manufacturing\ candidates\ in\ the\ frontier] \times P(entry|such\ a\ candidate)$$

For consistency with this accounting relationship, the conversion margin is measured as the pooled share of higher-complexity manufacturing candidate observations that subsequently enter. Table 4 reports the two margins by region; table 5 decomposes the regional gap.

[Insert Table 3 Here]

The availability margin is unambiguous. A Latin American country-year contained on average 0.49 such candidates in its frontier—one option roughly every second year—against 2.67 globally and 3.43 in East and Southeast Asia. At least one such candidate appeared in only 16.4% of Latin American country-years, against 38.1% globally and 51.9% in East and Southeast Asia; in roughly five out of six Latin American country-years, the peer-revealed opportunity set contained no higher-complexity manufacturing option at all.

[Insert Table 5 Here]

On the benchmark calculation, roughly four fifths of the East and Southeast Asia–Latin America gap in expected upgrading entries per country-year is attributable to the availability of upgrading candidates, and roughly one fifth to their realization. The compositional problem thus dominates the realization problem in the data, even within the narrow subset where a realization gap is visible—although the small number of Latin American observations in this category, itself a direct consequence of the scarcity, makes the realization component imprecise. Given the illustrative nature of the calculation, the robust statement is the ordering, not the exact split.

## 4.5. A frontier that did not upgrade — and heterogeneity within the region

Was the compositional gap a snapshot artifact or a trajectory? Because the frontier is redefined every year as export structures and peer groups evolve, its composition can drift—toward or away from upgrading content.

[Insert Figure 6 Here]

Figure 6 traces the manufacturing share of the frontier. In East and Southeast Asia, the share climbed steadily and remained high after 2000. Latin America's trajectory was different in shape, not merely in level: the manufacturing share rose appreciably during the 1970s and 1980s—in some periods approaching the global benchmark—but the movement was not consolidated, and after 2000 the share fell back. The frontier never durably reoriented toward manufacturing. The post-2000 decline coincides with the China-led commodity super-cycle, which shifted relative prices toward primary products and encouraged reprivatization across much of the region; part of the movement in figures 6–7 may therefore reflect price-driven changes in RCA structures—of Latin American countries and of their peers—rather than changes in underlying capabilities.

[Insert Figure 7 Here]

Figure 7 narrows the lens to above-average-complexity manufacturing and makes the contrast starker. From the late 1990s the East and Southeast Asian frontier came to include a visibly rising share of such products—the frontier upgraded, in composition as well as in quantity. The Latin American share remained low throughout the entire period. The region's frontier thus reflected, year after year, the experience of peers whose own portfolios contained little higher-complexity manufacturing; the reference set reproduced the structure rather than pointing beyond it. Two caveats bound this reading. First, frontier evolution is a diagnostic description, not an explanation: East and Southeast Asia's frontier upgraded in part because those economies themselves upgraded, so the trajectory partly restates the outcome. Second, regional averages mask well-known heterogeneity within Latin America: since the 1990s, Mexico and parts of Central America have integrated deeply into North American manufacturing value chains, while much of South America experienced commodity-driven reprimarization.

## 4.6. What peer experience adds beyond relatedness

A natural objection is that imitation density merely repackages product-space relatedness: economies with similar baskets hold related products, so peer portfolios might simply mirror internal proximity. The two measures are indeed correlated—Spearman correlations over the 1965–2014 baseline candidates are about 0.61 pooled and average 0.85 within country-years—but correlation at the level of full rankings does not imply agreement at the level of the short lists that matter for screening. Comparing the Top 20 sets under each measure within each country-year, an average of 48.0% of the products are common on both lists. Thus, although the two measures are strongly correlated over their full rankings, roughly half of each short list is selected by only one of the two criteria.

[Insert Table 6 Here]

Table 6 shows that the differences between the two short lists are associated with meaningful differences in subsequent entry. Candidates included in both Top-20 sets enter at the highest rate (33.2%). Candidates included only in the imitation-density Top-20—that is, products outside the country-year's Top-20 relatedness-density set but recurrent among its structural peers—enter at 27.8%, more than twice the 11.8% rate among candidates included in neither set and above the 24.6% rate among relatedness-only candidates. These descriptive differences indicate that imitation density identifies additional candidates not prioritized by the relatedness-density screen, and vice versa. The appropriate interpretation is therefore one of complementarity: imitation density adds information from structurally similar economies alongside, rather than in place of, the internal-capability benchmark provided by relatedness density.

## 5. Discussion

### 5.1. Rereading Latin America's diversification record

The results reframe a familiar narrative, within an explicitly bounded domain. On the evidence assembled here, Latin American countries did not display a broad inability to act on the diversification opportunities revealed by the experience of comparable economies; they converted those opportunities into new comparative advantage at rates close to the benchmark. What distinguished the region was the content of the opportunities themselves. The peer-revealed opportunity set was thin in manufacturing, nearly empty of higher-complexity manufacturing, and—unlike East and Southeast Asia's—did not thicken over time; the illustrative decomposition attributes the bulk of the regional gap in upgrading entries to this availability margin. To the extent that "missed opportunities" implies options visibly validated by peer experience and left unexploited, the diagnosis finds little support at the level of the frontier as a whole; few upgrading-oriented options were validated by the experience of current structural peers in the first place. This is emphatically not a claim that Latin America lacked all feasible upgrading opportunities: options outside the peer-revealed set—accessible through foreign investment, deliberate industrial policy, technology transfer or regional value chains—are invisible to our measure by construction.

This reading is congruent with the structuralist account of Latin American development in which a productive structure specialized in low-complexity activities tends to reproduce itself (Prebisch, 1950; Ocampo, 2005; ECLAC, 2012). The present results identify an external channel of that reproduction, complementary to the internal channel emphasized by the product space. Internally, a shallow basket offers few related stepping-stones (Hausmann and Klinger, 2007). Externally—and this is the article's additional shallow basket surrounds a country with structural peers whose own experience contains little upgrading to reference. The two channels compound: the same specialization pattern that limits what is nearby also limits what is observable in comparable economies.

Four interpretive boundaries must be respected. First, the relationships documented here are systematic associations, not causal effects; imitation density does not identify the mechanisms—policy emulation, firm learning, investment, supply chains—through which peer experience matters, and it confers no normative ranking on products. The low manufacturing content of Latin America's frontier is a diagnostic finding about the composition of peer-revealed options, not evidence of failure or inefficiency by any individual country. Second, imitation density is one signal of feasibility, not its measurement; all claims about the opportunity set are claims about the set as revealed by peer experience. Third, RCA-based entry does not measure domestic value added, employment or firm-level technological learning, so even realized entry is an imperfect proxy for the developmental content of diversification. This caveat cuts in both directions of our regional comparison: part of East and Southeast Asia's rise in higher-complexity exports reflects integration into global value chains and processing trade, so gross-export-based complexity measures may overstate domestic capability accumulation there; the comparison nonetheless retains its force, since Latin America's frontier lacked even these assembly-driven opportunities. Fourth, RCA structures respond to relative prices: the commodity super-cycle of the 2000s mechanically raised primary-product RCAs across the region and its peers, so part of the post-2000 frontier trajectory may be a price phenomenon rather than a capability phenomenon (see section 4.5).

## 5.2. Implications for opportunity-identification in industrial policy

For the growing practice of data-driven opportunity identification—product-space screenings, growth diagnostics, export road maps—the results carry a caution, a design suggestion and a policy conclusion.

Caution concerns predictive validity as a criterion of usefulness. Imitation density predicts entry strongly, and by that standard it is an excellent screening variable; yet in Latin America the very set it identifies best is compositionally unable to carry structural change. A screening tool can be simultaneously accurate and developmentally uninformative if its output composition goes unexamined. Opportunity-identification exercises should therefore report the sectoral and complexity composition of the candidate sets they produce—and its evolution—alongside predicted entry probabilities, treating a persistently low-complexity frontier as itself a diagnostic result about the position of the country, not as a menu to be executed.

The design suggestion is a two-stage screening architecture that keeps feasibility and developmental value analytically separate. A first stage assesses realizability—imitation density, relatedness density, existing export presence and RCA history, input and supply-chain access, and concrete peer precedents. A second stage assesses developmental value—technology intensity, product complexity, prospects for domestic value added, productivity and learning spillovers, forward and backward linkages, world demand prospects and alignment with the green and digital transitions. Policy priority then reflects both stages jointly rather than predicted entry probability alone. The complementarity documented in section 4.6 also implies that peer-experience screening deserves a standing place next to product-space tools within the first stage, particularly for surfacing options that are not close to the current basket yet demonstrably compatible with structures like it. A natural extension is to distinguish structural peers from aspirational peers—economies with similar current structures but somewhat higher complexity, or economies that resembled the focal country in the past and subsequently upgraded—and to compare the candidate sets each group reveals; current peers offer high short-run realizability with low average upgrading content, aspirational peers the reverse, and making that trade-off explicit is itself useful to policymakers.

The policy conclusion follows from the central finding. When both the internal signal (relatedness) and the external signal (peer experience) reveal few upgrading pathways—precisely the Latin American configuration documented here—empirical market signals and peer emulation are insufficient guides for structural transformation. The findings thus provide an evidence-based rationale for path-defying, mission-oriented productive development policies: deliberate state action—capability-building, investment attraction, public procurement, technology adoption and regional coordination—aimed not at executing the current peer-revealed frontier but at shifting its composition, since what tomorrow's peers reveal depends on whose experience a country's structure makes relevant. This orientation aligns with ECLAC's productive development agenda (ECLAC, 2012). The finding that higher-complexity candidates, where they did appear in Latin America, translated into entry at lower rates additionally suggests that scarce upgrading options may require more deliberate accompaniment—finance, standards, infrastructure, demand—than ordinary frontier products, although the small samples involved make this a hypothesis for closer study rather than a conclusion.

## 6. Conclusion

Did Latin American countries fail to seize upgrading opportunities, or were such opportunities scarce in their diversification set to begin with? Making the question empirical requires making an opportunity set observable and being precise about which one. We measured, for each country, product and year, the intensity with which structurally similar economies already held products the country did not yet export competitively, validated that this imitation density measure robustly predicts subsequent entry, and evaluated the resulting peer-revealed frontier on two margins—availability and realization.

The results do not imply that Latin America lacked all feasible upgrading opportunities, or that policy implementation was unimportant. They show instead that the opportunities most strongly validated by the experience of contemporaneous structural peers were disproportionately concentrated on low-complexity products. Within this peer-revealed set, the regional upgrading gap arose primarily through availability—Latin American frontiers contained few manufacturing options, almost no higher-complexity manufacturing options, and showed none of the compositional upgrading that the East and Southeast Asian frontier displayed from the 1990s onward—while a smaller and less precisely estimated realization gap remained among the rare higher-complexity manufacturing candidates. The formal decomposition quantifies this ordering, attributing the bulk of the regional gap in expected upgrading entries to the availability margin.

Beyond the regional diagnosis, the article leaves two portable results. Analytically, diversification research should distinguish between the availability and realization of opportunities because the two imply different diagnoses and policy responses. Methodologically, peer-country experience contains information about diversification possibilities that internal relatedness does not capture, and vice versa, so the two screens are complements. Extending the framework to aspirational-peer benchmarks, to the mechanisms behind the peer-experience signal, and to the question of what moves the composition of a country's revealed opportunity set are natural next steps.

## Bibliography

Abramovitz, M. (1986), "Catching up, forging ahead, and falling behind", The Journal of Economic History, vol. 46, No. 2, Cambridge, Cambridge University Press.

Bahar, D., R. Hausmann and C. A. Hidalgo (2014), "Neighbors and the evolution of the comparative advantage of nations: evidence of international knowledge diffusion?", Journal of International Economics, vol. 92, No. 1, Amsterdam, Elsevier.

Balassa, B. (1965), "Trade liberalisation and 'revealed' comparative advantage", The Manchester School, vol. 33, No. 2, Manchester, University of Manchester.

Balland, P.-A. et al. (2019), "Smart specialization policy in the European Union: relatedness, knowledge complexity and regional diversification", Regional Studies, vol. 53, No. 9, Abingdon, Taylor & Francis.

Basile, R., A. Parteka and R. Pittiglio (2018), "Export diversification and economic development: a dynamic spatial data analysis", Review of International Economics, vol. 26, No. 3, Hoboken, Wiley.

Boschma, R. (2017), "Relatedness as driver of regional diversification: a research agenda", Regional Studies, vol. 51, No. 3, Abingdon, Taylor & Francis.

Boschma, R., V. Martín and A. Minondo (2017), "Neighbour regions as the source of new industries", Papers in Regional Science, vol. 96, No. 2, Hoboken, Wiley.

Cadot, O., C. Carrère and V. Strauss-Kahn (2011), "Export diversification: what's behind the hump?", The Review of Economics and Statistics, vol. 93, No. 2, Cambridge, MA, MIT Press.

Cimoli, M. and G. Porcile (2014), "Technology, structural change and BOP-constrained growth: a structuralist toolbox", Cambridge Journal of Economics, vol. 38, No. 1, Oxford, Oxford University Press.

Coniglio, N. D. et al. (2021), "On the evolution of comparative advantage: path-dependent versus path-defying changes", Journal of International Economics, vol. 133, Amsterdam, Elsevier.

ECLAC (Economic Commission for Latin America and the Caribbean) (2012), Structural Change for Equality: An Integrated Approach to Development (LC/G.2524(SES.34/3)), Santiago.

Growth Lab at Harvard University (2026), The Atlas of Economic Complexity, Cambridge, MA, Harvard University.

Haraguchi, N., C. F. C. Cheng and E. Smeets (2017), "The importance of manufacturing in economic development: has this changed?", World Development, vol. 93, Amsterdam, Elsevier.

Hausmann, R., J. Hwang and D. Rodrik (2007), "What you export matters", Journal of Economic Growth, vol. 12, No. 1, New York, Springer.

Hausmann, R. and B. Klinger (2007), "The structure of the product space and the evolution of comparative advantage", CID Working Paper, No. 146, Cambridge, MA, Center for International Development, Harvard University.

Henn, C., C. Papageorgiou and N. Spatafora (2013), "Export quality in developing countries", IMF Working Paper, No. 13/108, Washington, D.C., International Monetary Fund.

Hidalgo, C. A. and R. Hausmann (2009), "The building blocks of economic complexity", Proceedings of the National Academy of Sciences, vol. 106, No. 26, Washington, D.C., National Academy of Sciences.

Hidalgo, C. A. et al. (2007), "The product space conditions the development of nations", Science, vol. 317, No. 5837, Washington, D.C., American Association for the Advancement of Science.

Hidalgo, C. A. et al. (2018), "The principle of relatedness", Unifying Themes in Complex Systems IX, A. J. Morales et al. (eds.), Cham, Springer.

Imbs, J. and R. Wacziarg (2003), "Stages of diversification", American Economic Review, vol. 93, No. 1, Nashville, American Economic Association.

Jeong, D., S. Lee and W. Eum (2024), "Learning from rivals: diversification paths from countries with similar export structures", SSRN Working Paper, No. 4850246, Rochester, NY, Social Science Research Network.

Jun, B. et al. (2020), "Bilateral relatedness: knowledge diffusion and the evolution of bilateral trade", Journal of Evolutionary Economics, vol. 30, New York, Springer.

Kim, L. (1997), Imitation to Innovation: The Dynamics of Korea's Technological Learning, Boston, Harvard Business School Press.

Klinger, B. and D. Lederman (2006), "Diversification, innovation, and imitation inside the global technological frontier", Policy Research Working Paper, No. 3872, Washington, D.C., World Bank.

Klinger, B. and D. Lederman (2011), "Export discoveries, diversification and barriers to entry", Economic Systems, vol. 35, No. 1, Amsterdam, Elsevier.

Lall, S., J. Weiss and J. Zhang (2006), "The 'sophistication' of exports: a new trade measure", World Development, vol. 34, No. 2, Amsterdam, Elsevier.

Lee, K. (2013), Schumpeterian Analysis of Economic Catch-up: Knowledge, Path-Creation, and the Middle-

Income Trap, Cambridge, Cambridge University Press.
Lin, J. Y. (2012), New Structural Economics: A Framework for Rethinking Development and Policy, Washington, D.C., World Bank.
Lin, J. Y. and Y. Fu (2024), "New Structural Economics and latent comparative advantage", New Structural Economics, J. Y. Lin (ed.), Cambridge, Cambridge University Press.
Lin, J. Y. and C. Monga (2011), "Growth identification and facilitation: the role of the state in the dynamics of structural change", Development Policy Review, vol. 29, No. 3, London, Overseas Development Institute.
Ocampo, J. A. (2005), "The quest for dynamic efficiency: structural dynamics and economic growth in developing countries", Beyond Reforms: Structural Dynamics and Macroeconomic Vulnerability, J. A. Ocampo (ed.), Stanford, Stanford University Press/ECLAC.
Pinheiro, F. L. et al. (2022), "The time and frequency of unrelated diversification", Research Policy, vol. 51, No. 8, Amsterdam, Elsevier.
Prebisch, R. (1950), The Economic Development of Latin America and Its Principal Problems (E/CN.12/89/Rev.1), New York, United Nations.
Rodrik, D. (2016), "Premature deindustrialization", Journal of Economic Growth, vol. 21, No. 1, New York, Springer.

**Acknowledgement**

This work is based on the first author's doctoral dissertation

**Conflicts of interest**

The authors declare that they have no conflict of interest.

# Figures

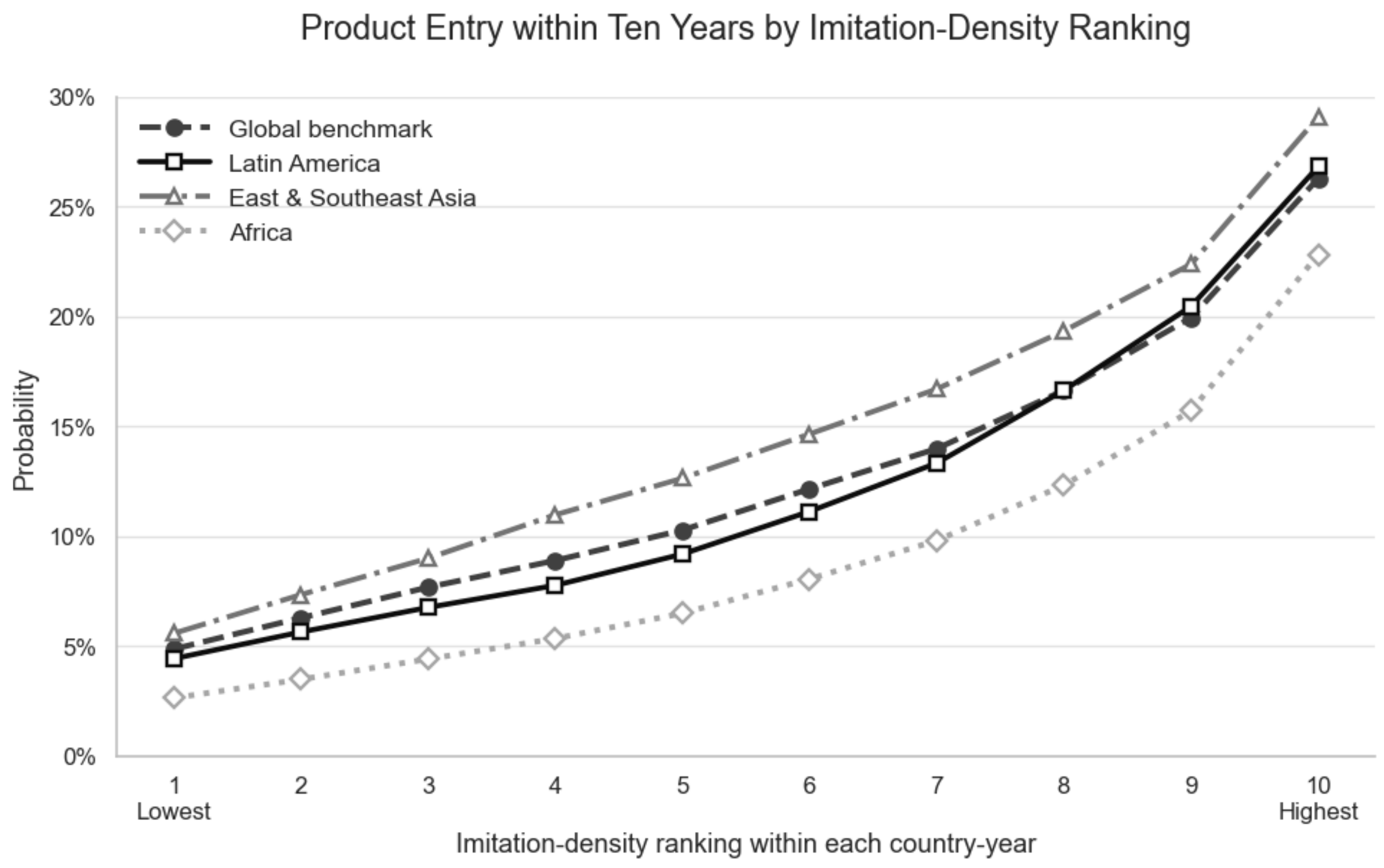

Figure 1. Product Entry within Ten Years by Imitation Density Ranking

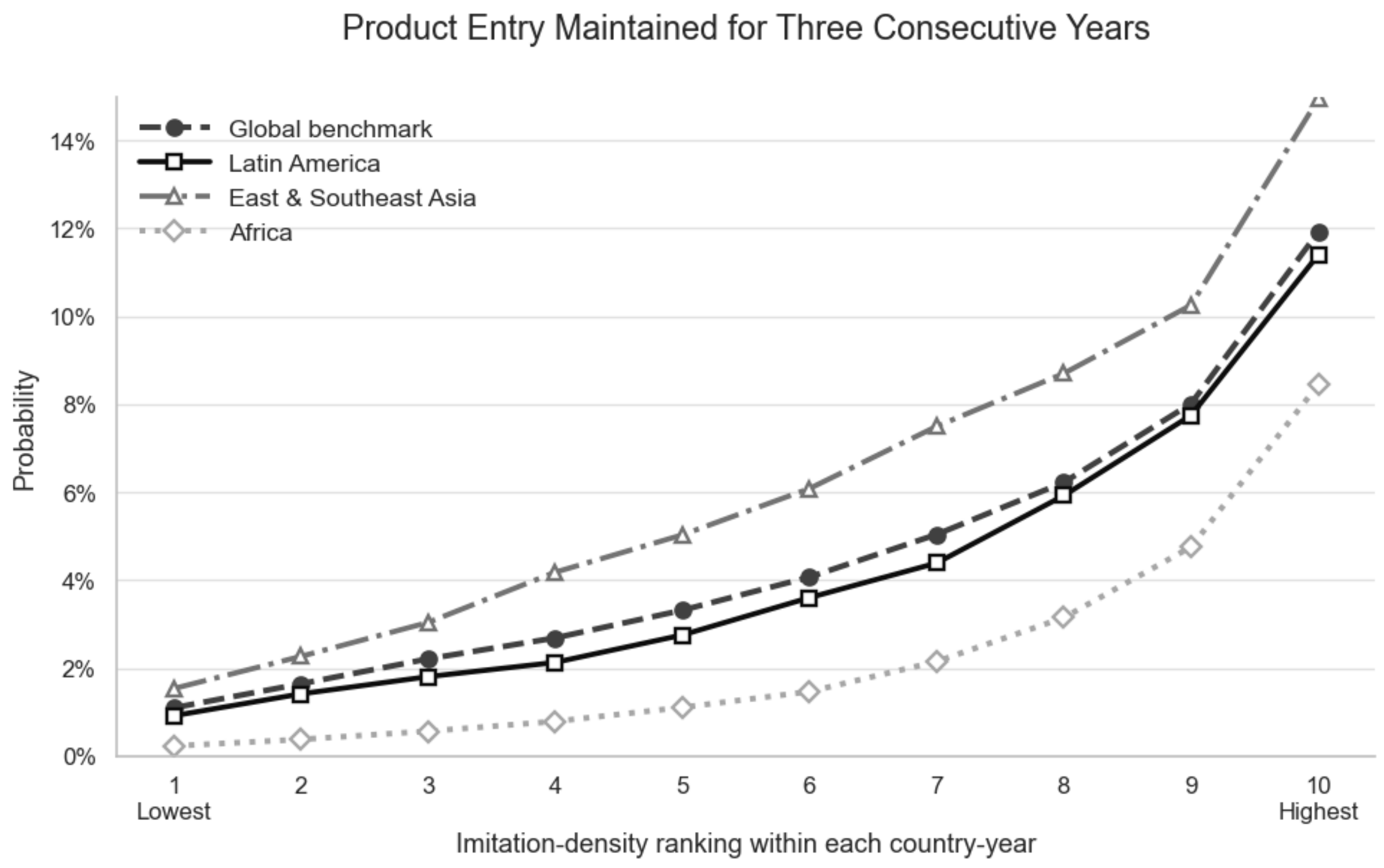

Figure 2 — Product entry maintained for three consecutive years, by imitation-density decile.

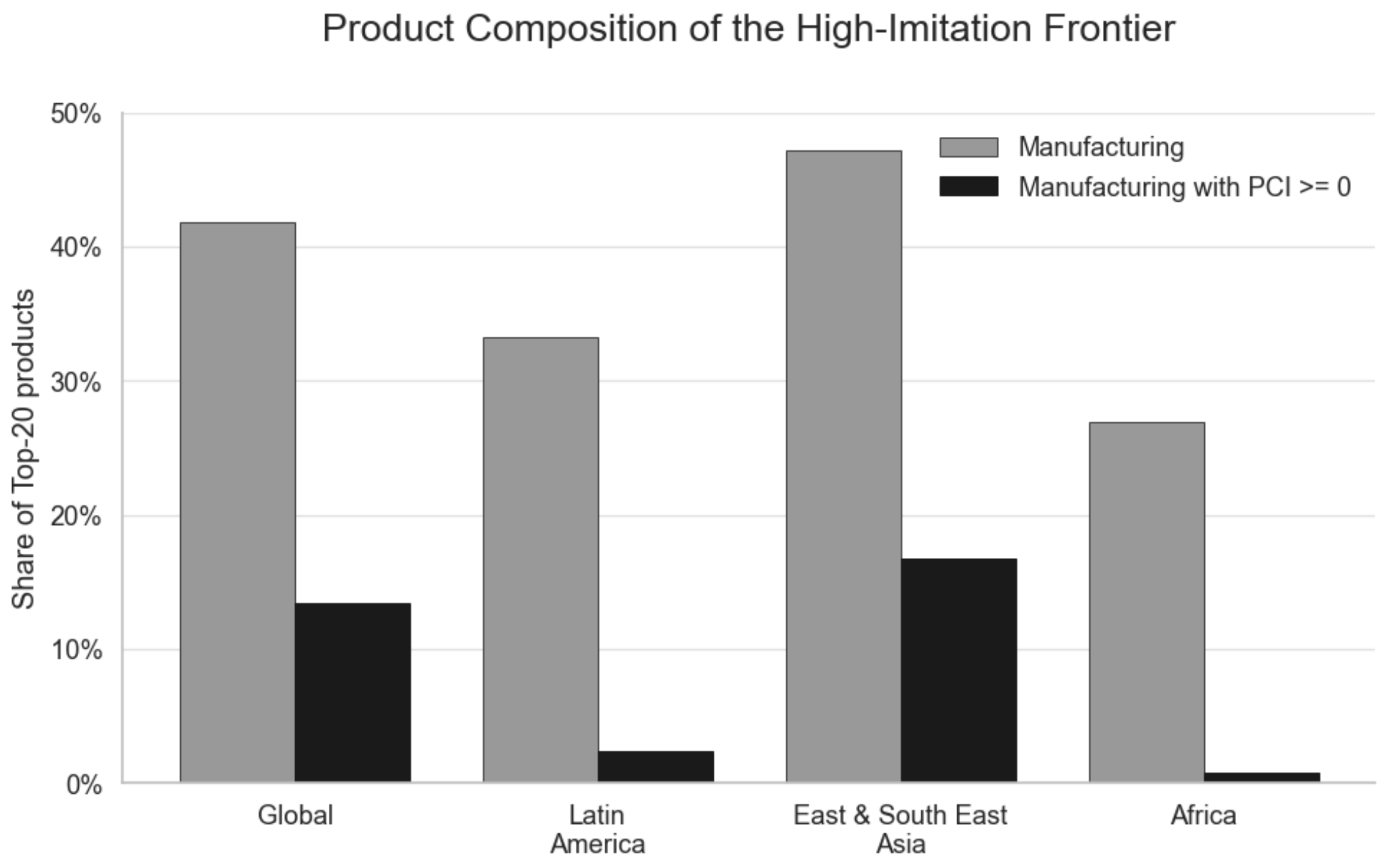

Figure 3 — Product composition of the peer-revealed frontier by region

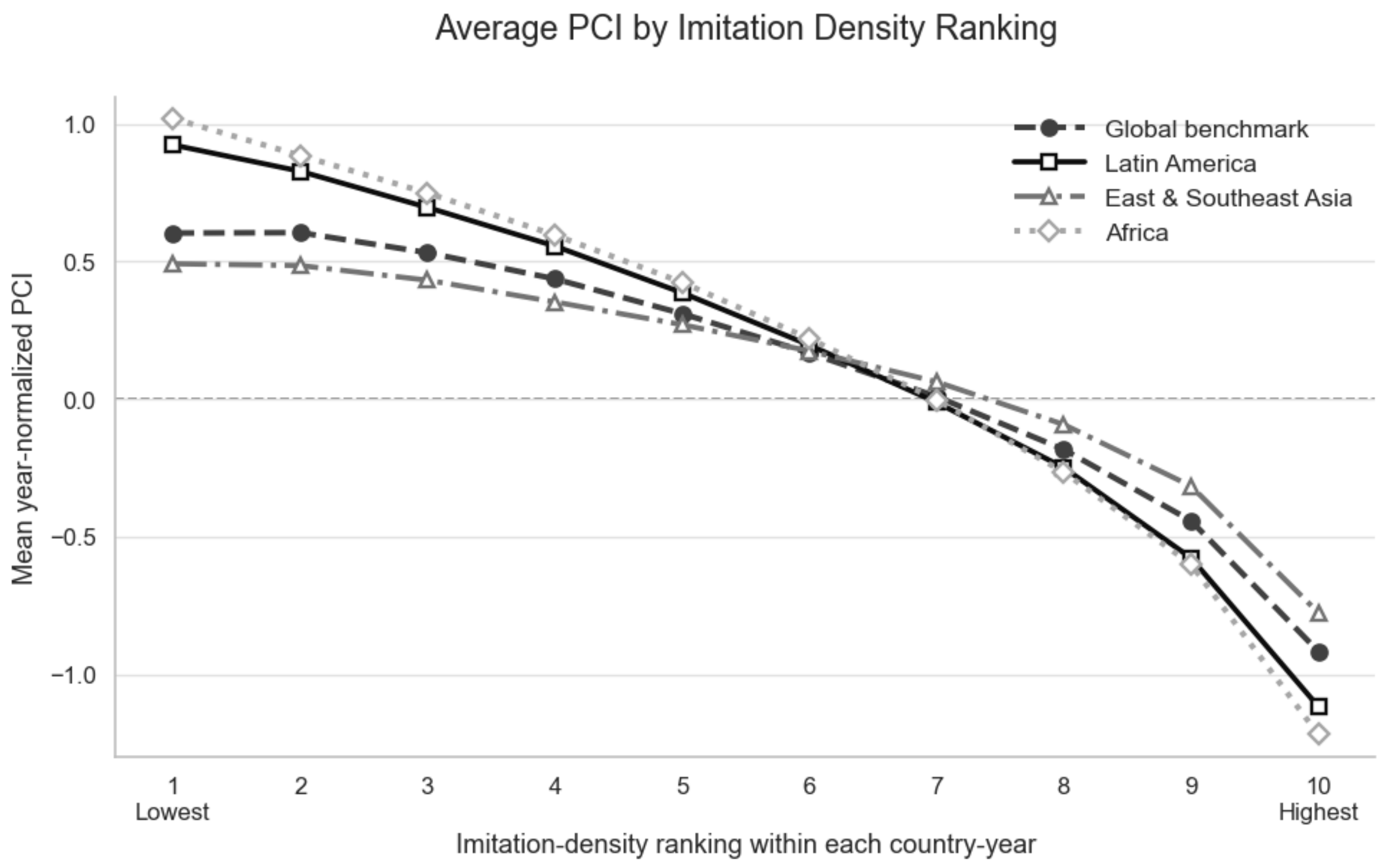

Figure 4 — Average standardized PCI by imitation-density decile

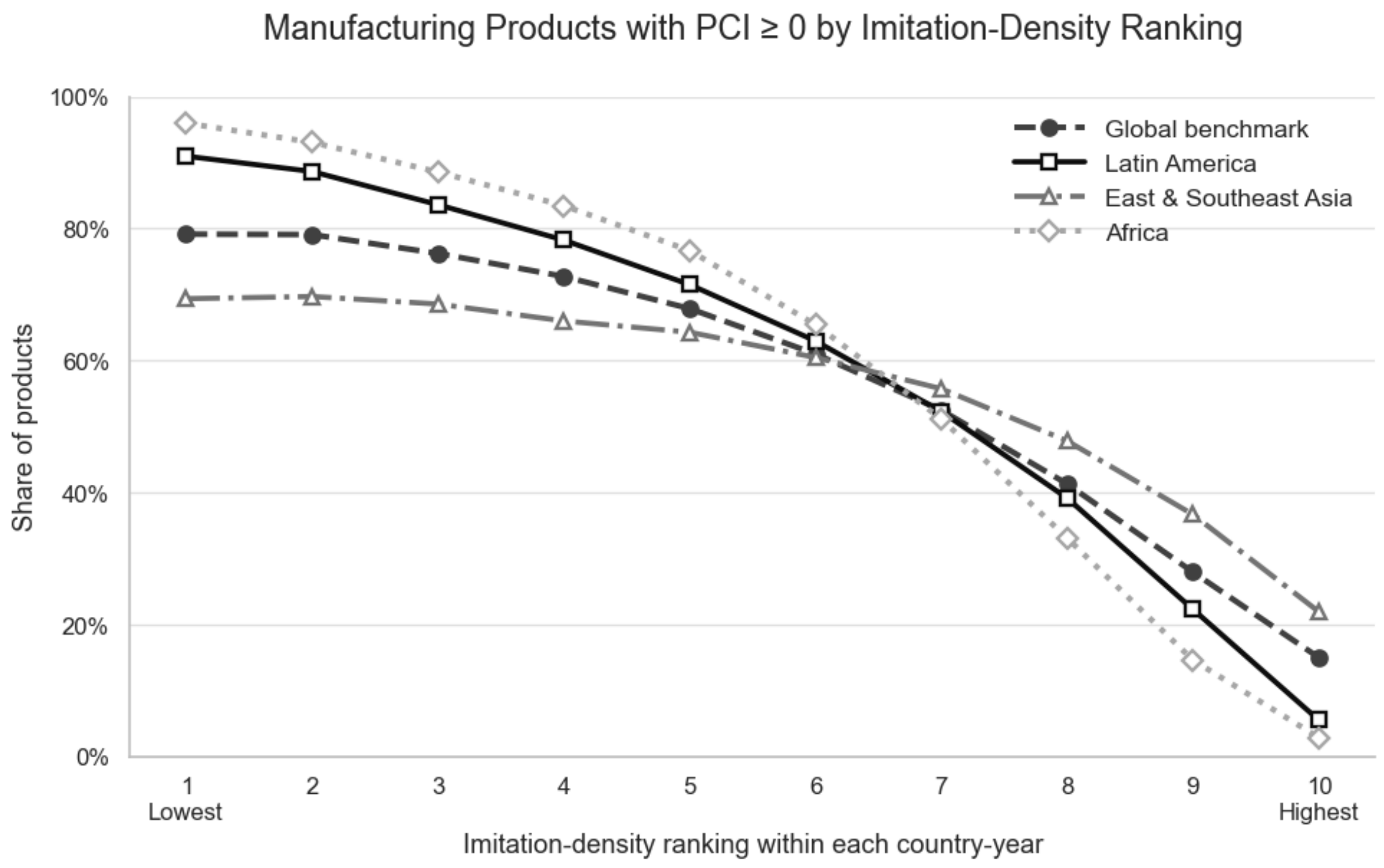

Figure 5 — Share of manufacturing products with PCI ≥ 0 by imitation-density decile.

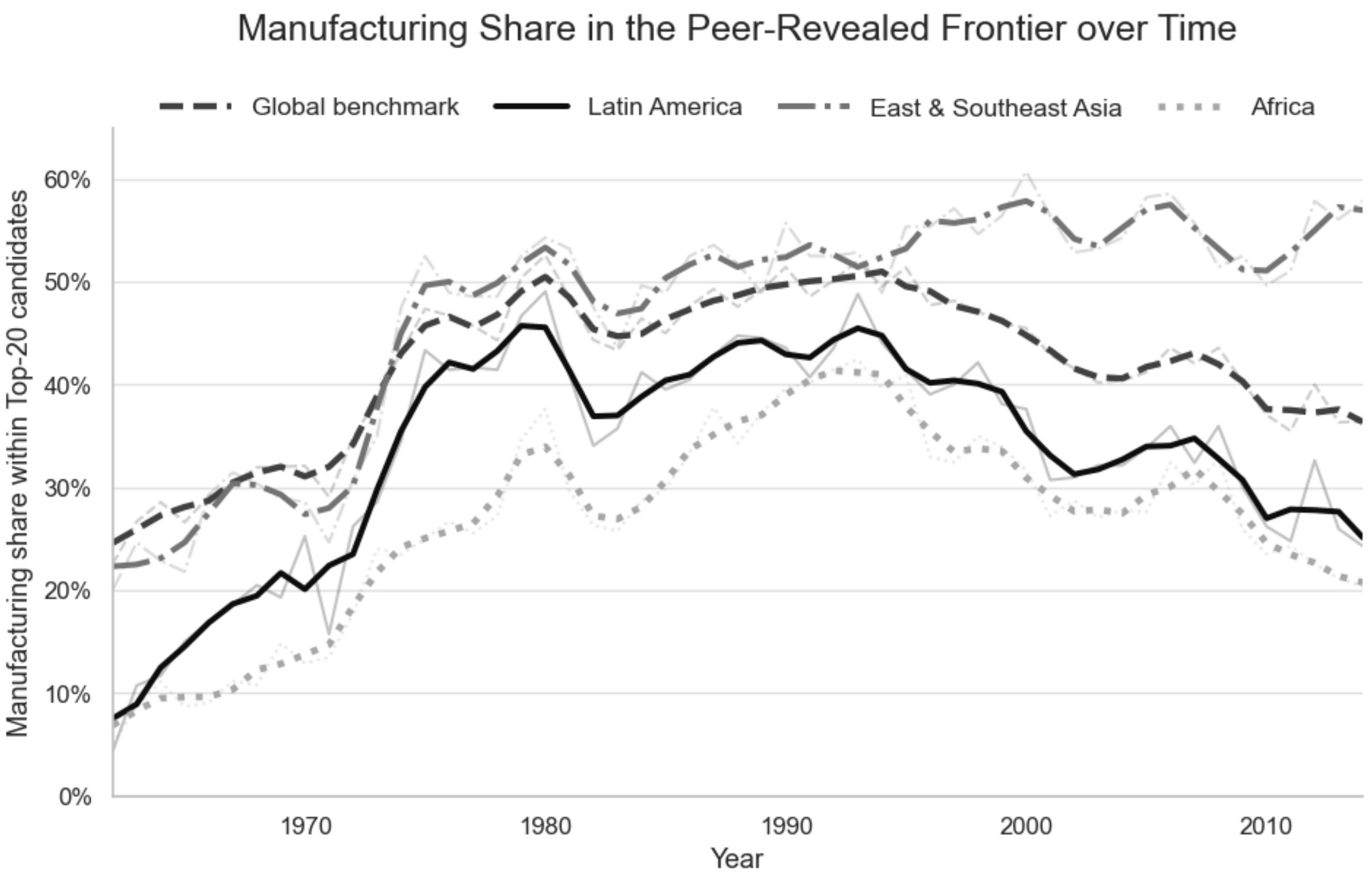

Figure 6 — Manufacturing share in the peer-revealed frontier over time, three-year moving averages.

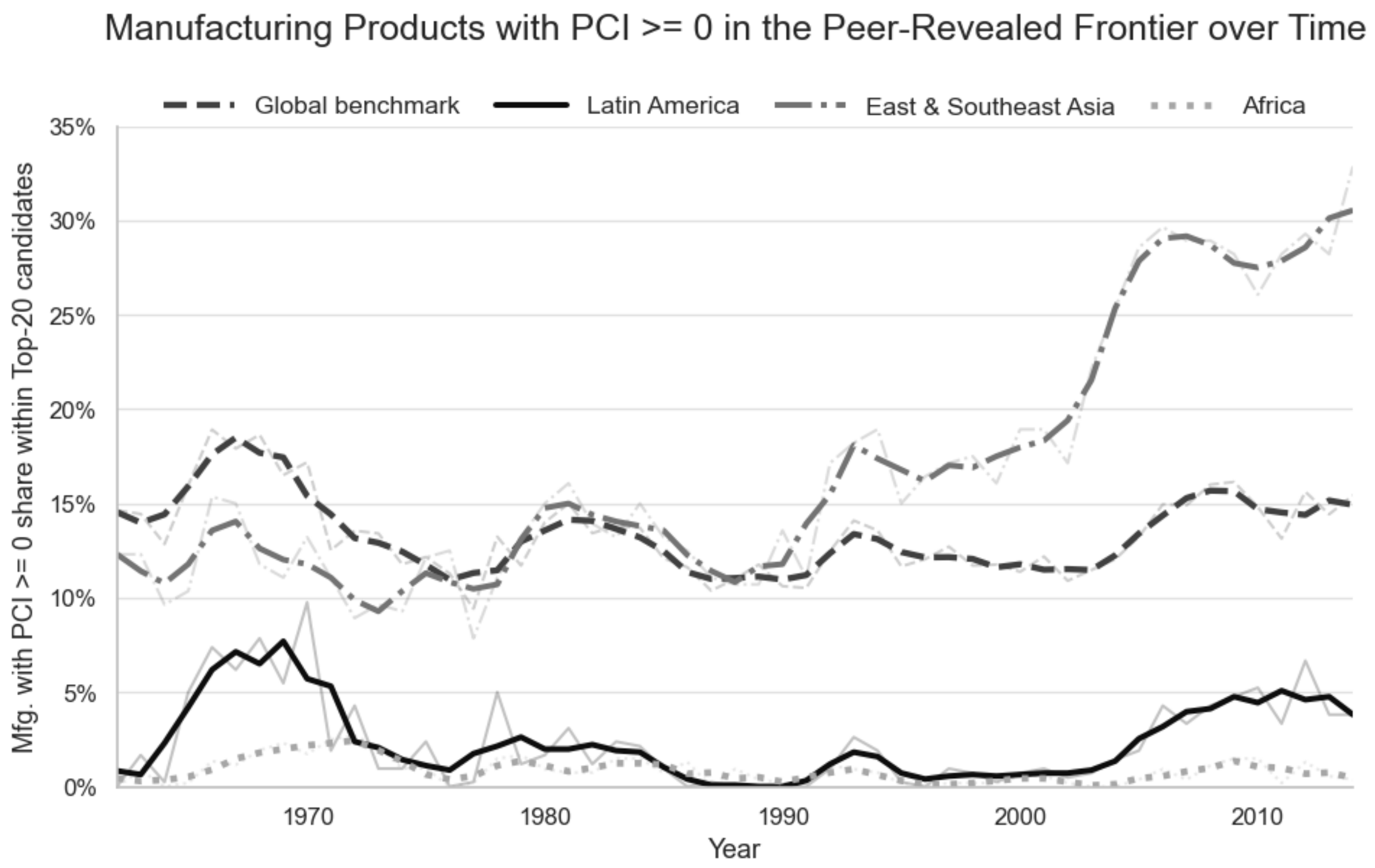

Figure 7 — Share of manufacturing products with PCI ≥ 0 in the peer-revealed frontier over time.

# Tables

### Table 1. Imitation density and product entry within ten years

| Dependent variable: Product Entry | (1) ID only | (2) RD only | (3) ID and RD | (4) Product controls | (5) Product FE |
|---|---|---|---|---|---|
| Imitation density × 10 | **0.0680***<br>(0.0028)** | | **0.0469***<br>(0.0029)** | **0.0487***<br>(0.0032)** | **0.0466***<br>(0.0035)** |
| Relatedness density × 10 | | **0.1849***<br>(0.0078)** | **0.0834***<br>(0.0069)** | **0.0885***<br>(0.0061)** | **0.0914***<br>(0.0069)** |
| Manufacturing product | | | | −0.0031<br>(0.0039) | |
| Year-normalized PCI | | | | **0.0077***<br>(0.0026)** | **0.0321***<br>(0.0037)** |
| Country-year fixed effects | Yes | Yes | Yes | Yes | Yes |
| Product fixed effects | No | No | No | No | Yes |
| Observations | 2,904,005 | 2,904,005 | 2,904,005 | 2,904,005 | 2,904,005 |

*Note: Dependent variable equals one when a candidate product becomes active within ten years of the base year. Imitation density and relatedness density are multiplied by 10; coefficients indicate the change in entry probability associated with a 0.10 increase in each measure. Standard errors in parentheses. *** $p < 0.01$.*

**Table 2. Imitation density and product entry maintained for three consecutive years**

| Dependent variable: Product entry for three consecutive years | (1) ID only | (2) RD only | (3) ID and RD | (4) Product controls | (5) Product FE |
|---|---|---|---|---|---|
| Imitation density × 10 | **0.0350***** **(0.0014)** | | **0.0258***** **(0.0017)** | **0.0283***** **(0.0017)** | **0.0245***** **(0.0018)** |
| Relatedness density × 10 | | **0.0925***** **(0.0031)** | **0.0367***** **(0.0039)** | **0.0463***** **(0.0035)** | **0.0528***** **(0.0037)** |
| Manufacturing product | | | | 0.0023 (0.0021) | |
| Year-normalized PCI | | | | **0.0107***** **(0.0014)** | **0.0226***** **(0.0024)** |
| Country-year fixed effects | Yes | Yes | Yes | Yes | Yes |
| Product fixed effects | No | No | No | No | Yes |
| Observations | 2,904,005 | 2,904,005 | 2,904,005 | 2,904,005 | 2,904,005 |

*Note: Dependent variable equals one when the candidate product is active in three consecutive years within the ten-year window. See notes to table 1.*

## Table 3. Regional characteristics of the peer-revealed frontier

| Region | Entry (%) | 3-year maintained entry (%) | Manufacturing (%) | Mfg. with PCI ≥ 0 (%) | Mean PCI |
|---|---|---|---|---|---|
| Global benchmark | 30.7 | 14.6 | 41.8 | 13.4 | −1.112 |
| Latin America | 30.7 | 13.7 | 33.2 | 2.4 | −1.378 |
| East and Southeast Asia | 31.9 | 16.9 | 47.2 | 16.8 | −1.021 |
| Africa | 28.6 | 11.3 | 26.9 | 0.8 | −1.563 |

*Note: Frontier defined as the twenty top-ranked candidates by imitation density in each country-year; regional figures average country-year values. Entry and maintained entry measured over the ten years following the base year. PCI standardized within years.*

**Table 4. Higher-complexity manufacturing candidates in the frontier: availability and entry**

| Region | Mean PCI ≥ 0 Mfg. candidates | Country-years with ≥1 (%) | Share in Top-20 (%) | Entry (%) | 3-year maintained entry (%) |
|---|---|---|---|---|---|
| Global benchmark | 2.67 | 38.1 | 13.4 | 32.9 | 19.4 |
| Latin America | 0.49 | 16.4 | 2.5 | 21.2 | 10.4 |
| East & Southeast Asia | 3.43 | 51.9 | 17.1 | 30.1 | 19.3 |
| Africa | 0.17 | 8.2 | 0.9 | 29.6 | 10.4 |

*Note: Higher-complexity manufacturing candidates are frontier products with Mfg = 1 and year-standardized PCI ≥ 0. Results cover 1965–2014. Availability measures are calculated at the country-year level, while entry and maintained-entry rates are pooled across candidate-product observations within each group. The global benchmark includes all sample countries.*

## Table 5. Decomposition of the regional gap in expected upgrading entries

| Comparison | Expected upgrading entries per country-year | Gap | Availability contribution | Realization contribution |
|---|---|---|---|---|
| East and Southeast Asia | 3.43 × 0.301 ≈ 1.03 | | | |
| Latin America | 0.49 × 0.212 ≈ 0.10 | | | |
| East/Southeast Asia − Latin America | | ≈ 0.93 | ≈ 0.75 (≈ 81%) | ≈ 0.17 (≈ 19%) |

*Note: Illustrative decomposition using the rounded descriptive statistics of table 4: the availability contribution is calculated as (3.43 − 0.49) × [(0.301 + 0.212)/2] and the realization component as (0.301 − 0.212) × [(3.43 + 0.49)/2].*

**Table 6. Product entry by imitation-density and relatedness-density top-ranked groups**

| Candidate group | Entry (%) | 3-year maintained entry (%) |
|---|---|---|
| Top-ranked in both measures | 33.2 | 15.1 |
| Top-ranked in imitation density only | 27.8 | 13.5 |
| Top-ranked in relatedness density only | 24.6 | 10.8 |
| In neither top-ranked group | 11.8 | 4.1 |

*Note: "Top-ranked" denotes inclusion in the respective Top-20 set within each country-year. Groups are defined over baseline candidate observations for 1965–2014. See notes to table 3.*